\documentclass[fleqn,usenatbib]{mnras}

\usepackage{newtxtext,newtxmath}
\usepackage{graphicx}

\usepackage[T1]{fontenc}

\DeclareRobustCommand{\VAN}[3]{#2}
\let\VANthebibliography\thebibliography
\def\thebibliography{\DeclareRobustCommand{\VAN}[3]{##3}\VANthebibliography}

\usepackage{graphicx}	
\usepackage{amsmath}	
\usepackage{scalerel}
\usepackage{tikz}
\usepackage{soul}

\usetikzlibrary{svg.path}

\soulregister\cite7
\soulregister\citet7
\soulregister\citep7
\soulregister\ref7
\soulregister\pageref7

\definecolor{orcidlogocol}{HTML}{A6CE39}
\tikzset{
  orcidlogo/.pic={
    \fill[orcidlogocol] svg{M256,128c0,70.7-57.3,128-128,128C57.3,256,0,198.7,0,128C0,57.3,57.3,0,128,0C198.7,0,256,57.3,256,128z};
    \fill[white] svg{M86.3,186.2H70.9V79.1h15.4v48.4V186.2z}
                 svg{M108.9,79.1h41.6c39.6,0,57,28.3,57,53.6c0,27.5-21.5,53.6-56.8,53.6h-41.8V79.1z M124.3,172.4h24.5c34.9,0,42.9-26.5,42.9-39.7c0-21.5-13.7-39.7-43.7-39.7h-23.7V172.4z}
                 svg{M88.7,56.8c0,5.5-4.5,10.1-10.1,10.1c-5.6,0-10.1-4.6-10.1-10.1c0-5.6,4.5-10.1,10.1-10.1C84.2,46.7,88.7,51.3,88.7,56.8z};
  }
}

\newcommand\orcidicon[1]{\href{https://orcid.org/#1}{\mbox{\scalerel*{
\begin{tikzpicture}[yscale=-1,transform shape]
\pic{orcidlogo};
\end{tikzpicture}
}{|}}}}

\title[Stability of mass-loaded galactic winds]{Highly-mass-loaded hot galactic winds are unstable to cool filament formation}

\author[Nguyen et al.]{Dustin D.~Nguyen $^{1,2,3}$\thanks{NASA FINESST Fellow, e-mail: dnguyen.phys@gmail.com} \orcidicon{0000-0002-1875-6522}, Todd A.~Thompson $^{1,2,3}$ \orcidicon{0000-0003-2377-9574}, Evan E.~Schneider$^{4,5}$ \orcidicon{0000-0001-9735-7484}, \& Ashley P. Tarrant$^{1,2,3}$ \orcidicon{0000-0002-9691-3116} \\
  $^{1}$Center for Cosmology and Astro-Particle Physics, Ohio State University, 140 W.~18th Ave, Columbus, OH 43210 \\
  $^{2}$Department of Physics, Ohio State University, 191 W. Woodruff Ave, Columbus, OH 43210 \\
  $^{3}$Department of Astronomy, Ohio State University, 140 W.~18th Ave, Columbus, OH 43210 \\
  $^{4}$Department of Physics and Astronomy, University of Pittsburgh, Pittsburgh, PA 15260, USA \\
  $^{5}$Pittsburgh Particle Physics, Astrophysics, and Cosmology Center (PITT PACC), University of Pittsburgh, Pittsburgh, PA 15260, USA}

\date{Accepted XXX. Received YYY; in original form ZZZ}

\pubyear{2023}

\begin{document}
\label{firstpage}
\pagerange{\pageref{firstpage}--\pageref{lastpage}}
\maketitle

\begin{abstract}
When cool clouds are ram-pressure accelerated by a hot supersonic galactic wind, some of the clouds may be shredded by hydrodynamical instabilities and incorporated into the hot flow. Recent one-dimensional steady-state calculations show how cool cloud entrainment directly affects the bulk thermodynamics, kinematics, and observational characteristics of the hot gas. In particular, mass-loading decelerates the hot flow and changes its entropy. Here, we investigate the stability of planar and spherical mass-loaded hot supersonic flows using both perturbation analysis and three-dimensional time-dependent radiative hydrodynamical simulations. We show that mass-loading is stable over a broad range of parameters and that the 1D time-steady analytic solutions exactly reproduce the 3D time-dependent calculations, provided that the flow does not decelerate sufficiently to become subsonic. For higher values of the mass-loading, the flow develops a sonic point and becomes thermally unstable, rapidly cooling and forming elongated dense cometary filaments. We explore the mass-loading parameters required to reach a sonic point and the radiative formation of these filaments. For certain approximations, we can derive simple analytic criteria. In general a mass-loading rate similar to the initial mass outflow rate is required. In this sense, the destruction of small cool clouds by a hot flow may ultimately spontaneously generate fast cool filaments, as observed in starburst superwinds. Lastly, we find that the kinematics of filaments is sensitive to the slope of the mass-loading function. Filaments move faster than the surrounding wind if mass-loading is over long distances whereas filaments move slower than their surroundings if mass-loading is abrupt. 
 
\end{abstract}




\section{Introduction}
Galactic outflows are a critical component of galaxy evolution. Winds are commonly found in star-forming galaxies at both low- and high-redshift \citep{Martin1998,Pettini2001,Rubin2010,Heckman2015,Heckman2016}. They modulate star formation \citep{Larson1974,Heckman1990}, shape the stellar mass and mass-metallicity relations \citep{Dekel1986,Finlator2008,Peeples2011}, and advect metals into the circumgalactic and intergalactic medium \citep{Aguirre2001,Scannapieco2002,Tremonti2004,Oppenheimer2006,Oppenheimer2008}. Evidence for high velocity cool neutral and ionized gas is observed in local starburst galaxies and high-redshift star-forming galaxies in both emission and absorption (e.g., \citealt{Veilleux2005,Heckman2017} and references therein).


A possible explanation for the origin of high velocity cool material seen in observations is that cool clouds are ram-pressure accelerated by a hot wind. However, low-column density cool clouds immersed in a hot supersonic wind will be shredded by hydrodynamic instabilities and mixed into the hot outflow before they can be accelerated (see \citealt{Zhang2017} for a review). The acceleration time, $t_\mathrm{drag}$, can be written in terms of destruction time (the ``cloud crushing" timescale), $t_\mathrm{cc}$, as $t_\mathrm{drag} \sim \sqrt{\chi} \, t_\mathrm{cc}$, where $\chi \gg 1 $ is the relative over-density of the cool cloud relative to the hot wind, implying that cool clouds should be destroyed before they are accelerated \citep{Klein1994}. While this result has been verified in a wide range of simulations  \citep{Klein1994,Cooper2009,Scannapieco2015,Zhang2017,Schneider2017,Li2020}, recent work shows that for sufficiently high column densities clouds can survive and even grow through mixing at the wind-cloud interface \citep{Gronke2018,Gronke2020}. 

The destruction and entrainment of low column density cool clouds into the hot superwind fluid has direct consequences for the bulk thermodynamics, kinematics, and X-ray emissivity of the hot wind \citep{Suchkov1996,Smith1996,Nguyen2021,Fielding2022}. Mass-loading ($\dot{\mu} \, \mathrm{[M_\odot \, yr^{-1} \, kpc^{-3}]})$ increases both the hot gas temperature and its density, while decelerating the flow. Further, mass-loading increases the hot gas entropy if the gas Mach number is above a critical value of $\mathcal{M} = \sqrt{(11+\sqrt(46))/5} \sim 1.9$ (see \citealt{Nguyen2021}) and decreases the entropy otherwise. Sufficiently strong mass-loading can decelerate a supersonic flow sufficiently to produce a sonic point (see below). 


A critical question is whether the 1D steady-state models are stable to mass-loading, even in the idealized cases where mass-loading is treated as a simple volumetric source of cool matter. \citet{Smith1996} suggested a smooth transition from super- to sub-sonic flow is not possible and that a shock is expected to develop, which could potentially cause a breakdown of the time-steady solutions. \citet{Schekinov1996} also argues that flows will become unstable near the sonic point, and that the instability growth rate may scale as the Mach number squared. 

Here, we provide a set of analytic investigations and numerical tests meant to illuminate the stability of mass-loaded hot outflows. This is an important step in critically assessing 1D steady-state models that purport to explain aspects of the hot gas profiles \citep{Suchkov1996,Nguyen2021} and 1D steady-state models that explicitly couple a distribution of cool clouds (with acceleration, deceleration, and growth) to the hot gas dynamics \citep{Cowie1981,Fielding2022}. We wish to answer the question of whether or not mass-loaded hot flows of the type typically invoked to explain the hot gas in M82 and other local starbursts are stable or unstable, and to understand how the bulk dynamics are effected, with direct relevance for observations.


In \autoref{sec:linear_stability}, we study the linear stability of mass-loaded flows. In \autoref{sec:Mload_required}, we quantify how much mass-loading is required to decelerate a steady supersonic wind through a sonic point. In \autoref{sec:Chollas}, we make a direct comparison between our semi-analytic 1D models and 3D hydrodynamical simulations and find that mass-loaded flows are stable provided they do not cross a sonic point. For high enough mass-loading, we find flows become thermally unstable and form dense filaments. We then discuss implications for galactic winds and multi-phase microphysics in \autoref{sec:discussion}. Lastly we summarize in \autoref{sec:conclusion} and highlight future directions.

\begin{figure}
    \centering
    \includegraphics[width=\columnwidth]{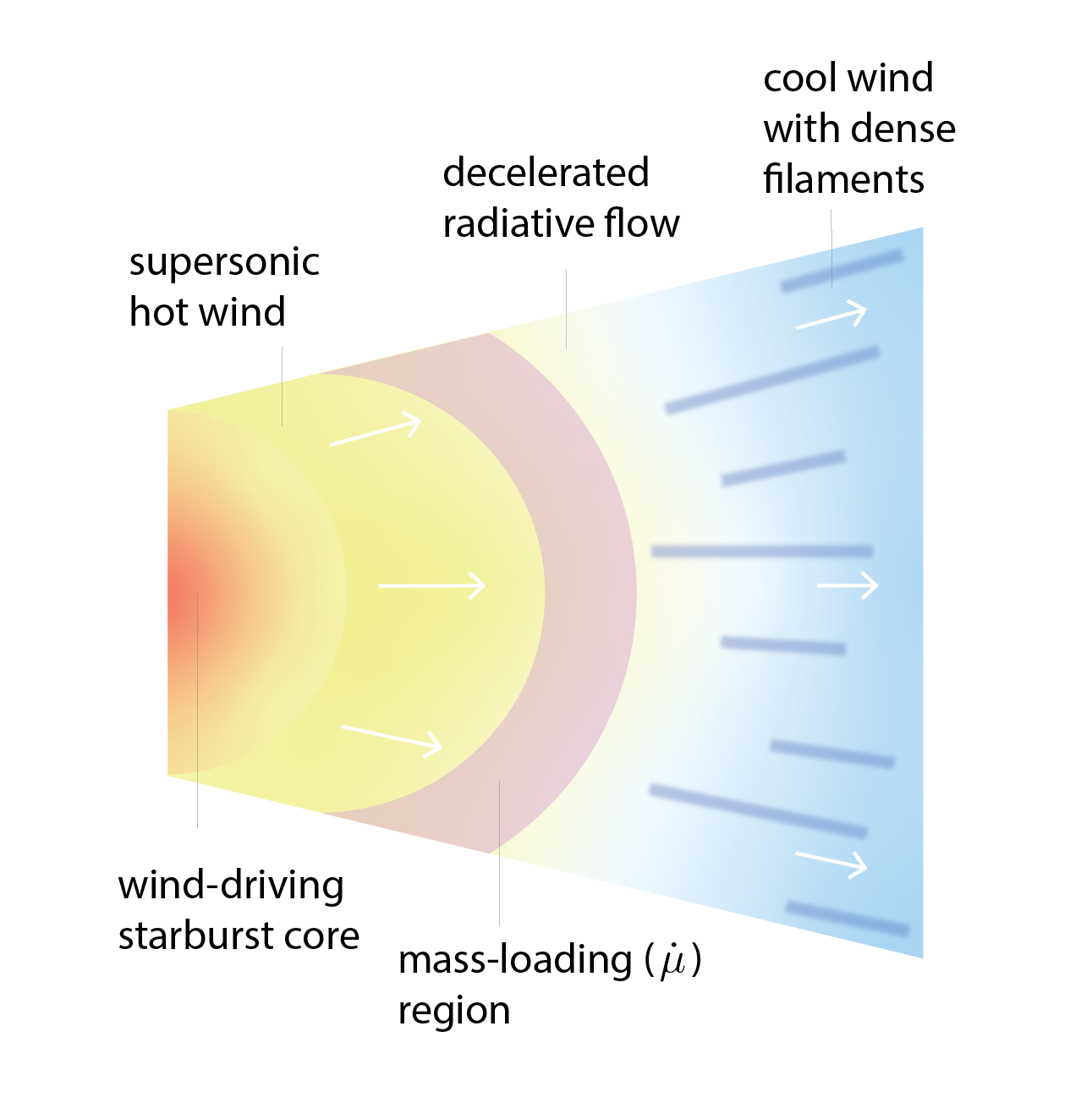}
    \caption{Schematic detailing the mass-loaded galactic wind. The wind emerges from the wind-driving region as a hot supersonic outflow and is then is mass-loaded from cool cloud entrainment. If the flow is sweeps up enough material, it will rapidly decelerate and become susceptible to rapid cooling and thermal instability, forming high velocity cool dense filaments.}
    \label{fig:doodle}
\end{figure}

\section{linear stability of mass-loaded flows} 
\label{sec:linear_stability}

Previous works have used steady-state models to quantify how mass entrainment impacts hot outflows \citep{Nguyen2021,Fielding2022}. Here, we consider the stability of mass-loaded winds for planar $(\hat{x})$, spherical $(\hat{r})$, and arbitrary area $A(x)$ flow tubes ($\hat{x}$). Throughout the analytics below, we will often take the high Mach number limit of a solution. We consider mass-loading to occur after the supersonic flow reaches its asymptotic bulk velocity, which occurs at around $\mathcal{M} \gtrsim 2$ for a launched spherical wind. The physical picture we have in mind is illustrated in Figure \ref{fig:doodle}.
     
\label{sec:mudot_analytics}
\subsection{Planar Mass-loaded Flow}
\label{sec:planar_mudot}
\begin{figure}
    \centering
    \includegraphics[width=\columnwidth]{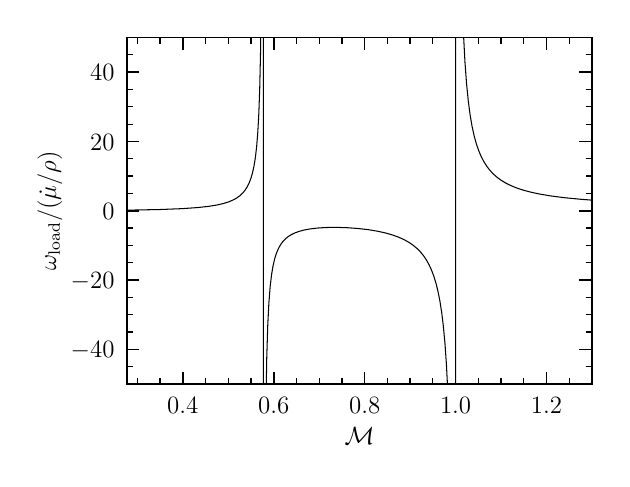}
    \caption{The instability growth rate (Eq.~\ref{eq:planar_growthrate}) divided by the mass-loading rate $\dot{\mu}/\rho$. The instability grows when $\omega_\mathrm{load}/\dot{\mu}/\rho>0$ and is damped when $\omega_\mathrm{load}/\dot{\mu}/\rho<0$. When being decelerated through critical point $\mathcal{M}_\mathrm{crit}=1$, the flow is shortly unstable, but is then becomes stabilized when is in the subsonic regime. Similarly, when being accelerated through $\mathcal{M}_\mathrm{crit}=1/\sqrt{3} \simeq 0.57$, the flow is unstable, but then becomes stable. The derived growth rates indicate that flows that pass through critical points $\mathcal{M}_\mathrm{crit}=1\,\mathrm{and}\,1/\sqrt{3}$ are unstable.}
    \label{fig:growth_rates}
\end{figure}

For a one-dimensional planar flow with mass-loading ($\dot{\mu}$ in g cm$^{-3}$ s$^{-1}$), the hydrodynamic equations are 
\begin{align}
    & \frac{\partial \rho}{\partial t} + \frac{\partial}{\partial x} ( \rho v ) = \dot{\mu}, \label{eq:td_planar_continuity} \\
    & \frac{\partial v}{\partial t} + v \frac{\partial v}{\partial x} = - \frac{1}{\rho} \frac{\partial P}{\partial x} - \frac{ v \dot{\mu}}{\rho}, \label{eq:td_planar_momentum} \\ 
    & \frac{\partial P}{\partial t} + v \frac{\partial P}{\partial x} = \frac{\dot{\mu} P}{\rho} \bigg( \frac{1}{2}(\gamma -1) \frac{\rho v^2}{P} - \gamma \bigg) + \frac{\gamma P}{\rho} \bigg( \frac{\partial \rho}{\partial t} + v \frac{\partial \rho}{\partial x} \bigg). \label{eq:td_planar_pressure}
\end{align}
The steady-state equations are 
\begin{align}
    & \frac{\partial }{\partial x} ( \rho_0 v_0 ) = \dot{\mu}, \label{eq:ss_planar_continuity} \\ 
    & v_0 \frac{\partial v_0 }{\partial x} = - \frac{1}{\rho_0} \frac{\partial P_0}{\partial x} - \frac{v_0 \dot{\mu}}{\rho_0}, \label{eq:ss_planar_momentum} \\ 
    & v_0 \frac{\partial P_0 }{\partial x} = \frac{\dot{\mu} P_0}{\rho_0 } \bigg( \frac{1}{2} (\gamma -1 ) \frac{\rho_0 v_0^2}{P_0} - \gamma \bigg) + \frac{\gamma v_0 P_0}{\rho_0} \frac{\partial \rho_0}{\partial x}. \label{eq:ss_planar_pressure}
\end{align}
We test the linear stability of the planar flow using perturbation analysis. Our volumetric mass-loading rate is a function of distance, so only the velocity, density, and pressure are perturbed as
\begin{align}
    v (x,t) &= v(x) + \delta v \exp [ - i (\omega t - k x ) ],
    \label{eq:perturbation} \\ 
    \rho (x,t) &= \rho(x) + \delta \rho \exp [ - i (\omega t - k x ) ], \nonumber \\
    P (x,t) &= P(x) + \delta P \exp [ - i (\omega t - k x ) ]. \nonumber  
    \label{eq:normal_modes}
\end{align}
Substituting the above perturbations into Equations \ref{eq:td_planar_continuity}, \ref{eq:td_planar_momentum}, and \ref{eq:td_planar_pressure}, we derive a dispersion relation for $\omega$ (see Appendix~\ref{sec:planar_disp_relation} for full dispersion relation).  
The system is unstable if $\omega$ is positive and complex, and stable otherwise. Only terms linear in the perturbations $\delta v$, $\delta \rho$, and $\delta P$ are kept. In the high wave number limit ($k \rightarrow \infty$) the dispersion relation is  
\begin{align}
    & k (1-\mathcal{M}^2)^2 v \rho ^2 + i \mathcal{M}^2 (1+ \gamma) ( 1- \gamma \mathcal{M}^2) \dot{\mu} \rho /2 \\  &  \hspace{4cm} - (1- 3 \mathcal{M}^2 (1 - \mathcal{M}^2) \rho ^2 \omega = 0. \nonumber
\end{align}
This high-$k$ limit of the dispersion relation is linear in $\omega$. The imaginary component of the frequency (i.e., growth rate), $\omega_\mathrm{load}$, is 
\begin{equation}
    \omega_\mathrm{load} = \frac{(1+\gamma)}{2} \frac{ \dot{\mu}}{\rho} \frac{(\gamma \mathcal{M}^4 + \mathcal{M}^2)}{(3 \mathcal{M}^4 - 4 \mathcal{M}^2 + 1 )},
    \label{eq:planar_growthrate}
\end{equation}
which has poles at
\begin{equation}
    \mathcal{M}_\mathrm{crit} = 1 / \sqrt{3} \ \mathrm{and} \ 1.
    \label{eq:Mach_crit}
\end{equation} The critical points $\mathcal{M}_\mathrm{crit}$ correspond to locations where instability growth and damping rates peak and is independent of $\gamma$. In Figure \ref{fig:growth_rates}, we plot $\omega_\mathrm{load}/(\dot{\mu}/\rho)$ as a function of Mach number (black solid line). We see that the instability grows as the flow is decelerated through the sonic point from $\mathcal{M} > 1$. However, immediately after transitioning the sonic point, the growth rate is negative indicating damping. For $\gamma=5/3$, in the highly supersonic limit $\mathcal{M} \gg 1$, the growth rate is 
\begin{equation}
    \omega_\mathrm{load} \simeq \frac{20}{27} \frac{\dot{\mu}}{\rho} \quad (\mathcal{M} \gg 1).
    \label{eq:planar_growthrate_highmach}
\end{equation}
Equation~\ref{eq:planar_growthrate_highmach} disagrees with the growth rate derived by \cite{Schekinov1996}, who found that $\omega_\mathrm{load} \propto \mathcal{M}^2$ for $\mathcal{M} \gg 1$. However, in their study the volumetric mass-loading rate was a function of temperature $\dot{\mu} = \dot{\mu}(T)$ which differs from our definition of mass-loading $\dot{\mu}=\dot{\mu}(r)$. For their variable $\alpha = \partial \log(\dot{\mu}) / \log (T) = 0$, our governing hydrodynamic equations are identical. In Appendix \ref{app:Schekinov}, we discuss possible reasons for the discrepancy. 
\subsection{Spherical Mass-loaded Flow} 
\label{sec:spherical_mudot}
For a spherical flow, the hydrodynamic equations are
\begin{align}
    & \frac{\partial \rho}{\partial t} + \frac{\partial}{\partial r^2} ( \rho v r^2 ) = \dot{\mu}, \label{eq:td_sph_continuity} \\
    & \frac{\partial v}{\partial t} + v \frac{\partial v}{\partial r} = - \frac{1}{\rho} \frac{\partial P}{\partial r} - \frac{ v \dot{\mu}}{\rho}, \label{eq:td_sph_momentum} \\ 
    & \frac{\partial P}{\partial t} + v \frac{\partial P}{\partial r} = \frac{\dot{\mu} P}{\rho} \bigg( \frac{1}{2}(\gamma -1) \frac{\rho v^2}{P} - \gamma \bigg) + \frac{\gamma P}{\rho} \bigg( \frac{\partial \rho}{\partial t} + v \frac{\partial \rho}{\partial r} \bigg). \label{eq:td_sph_pressure}
\end{align}
\begin{equation}
    \omega_\mathrm{load} = \frac{\dot{\mu}}{\rho} \frac{(\gamma+1)}{2} \frac{\gamma \mathcal{M}^4 + \mathcal{M}^2}{3 \mathcal{M}^4 - 4 \mathcal{M}^2 +1} - \frac{v}{r} \frac{[2(\gamma+1)\mathcal{M}^4 - 2 \mathcal{M}^2 +2 ]}{3 \mathcal{M}^4 -4 \mathcal{M}^2+1}.
    \label{eq:spherical_growthrate}
\end{equation}
The derived growth carries the same frequency derived in the planar case (Eq.~\ref{eq:planar_growthrate}), however there is an additional damping term that arises from spherical expansion, as expected. As in \citet{Nguyen2021}, we will consider mass-loading outside of the wind-driving region. In the high Mach number limit the growth rate is
\begin{equation}
    \omega_\mathrm{load} \simeq \frac{20}{27} \frac{\dot{\mu}}{\rho} - \frac{16}{9} \frac{v}{r} \quad (\mathcal{M} \gg 1 ) 
\end{equation}

\subsection{Non-spherical Mass-loaded Flow}
\label{sec:nonsph_mudot}
For a time-steady one-dimensional flow undergoing expansion into area element $A(x)$, the hydrodynamic equations are
\begin{align}
    & \frac{1}{A} \frac{\partial }{\partial x} ( A \rho_0 v_0 ) = \dot{\mu}, \\ 
    & v_0 \frac{\partial v_0 }{\partial x} = - \frac{1}{\rho_0} \frac{\partial P_0}{\partial x} - \frac{v_0 \dot{\mu}}{\rho_0}, \\ 
    & v_0 \frac{\partial P_0 }{\partial x} = \frac{\dot{\mu} P_0}{\rho_0 } \bigg( \frac{1}{2} (\gamma -1 ) \frac{\rho_0 v_0^2}{P_0} - \gamma \bigg) + \frac{\gamma v_0 P_0}{\rho_0} \frac{\partial \rho_0}{\partial x}.
\end{align}
Following the same procedures as described above in Sections \ref{sec:planar_mudot} and \ref{sec:spherical_mudot}, the derived growth rate in the high-$k$ limit is 
\begin{eqnarray}
    \omega_\mathrm{load} &=& \frac{\dot{\mu}}{\rho} \frac{(\gamma+1)}{2} \frac{\gamma \mathcal{M}^4 + \mathcal{M}^2}{3 \mathcal{M}^4 - 4 \mathcal{M}^2 +1} \noindent \\
    &-& 2v \frac{d \ln A}{dx} \frac{[2(\gamma+1)\mathcal{M}^4 - 2 \mathcal{M}^2 +2 ]}{3 \mathcal{M}^4 -4 \mathcal{M}^2+1}.
    \label{eq:nonsph_growthrate}
\end{eqnarray}
For $A \propto x^0$, we recover the growth rate of the planar flow (Eq.~\ref{eq:planar_growthrate}) and when $x\rightarrow r$ and $A \propto r^2$, we recover the growth rate of the spherical flow (Eq.~\ref{eq:spherical_growthrate}). In the high Mach number limit the growth rate is 
\begin{equation}
    \omega_\mathrm{load} \simeq \frac{20}{27} \frac{\dot{\mu}}{\rho} - \frac{32}{9} v \frac{d \ln A}{dx}.
\end{equation}

\section{Required Mass-loading For Transonic Transition}  
\label{sec:Mload_required}
In the previous section we found that the instability growth rates peak at $\mathcal{M}_\mathrm{crit} = 1 \ \mathrm{and} \ 1/\sqrt{3}$. We now derive the amount of mass-loading required for a supersonic flow to be decelerated through the sonic point. 
First, we expand the derivative of the Mach number $d\mathcal{M}/dr$ as: 
\begin{equation}
    \frac{d\mathcal{M}}{dr} = \frac{1}{(\gamma P / \rho )^{1/2}} \frac{dv}{dr} + \frac{v}{(\gamma P / \rho )^{1/2}} \frac{1}{2 \rho} \frac{d \rho}{dr} - \frac{v}{(\gamma P / \rho )^{1/2}} \frac{1}{2 P} \frac{dP}{dr}. 
    \label{eq:dMdr}
\end{equation}
From the steady-state equations (Eqs.~\ref{eq:td_sph_continuity}, \ref{eq:td_sph_momentum}, and \ref{eq:td_sph_pressure}, with $\partial/\partial t = 0$), we solve the three equations for the derivatives $dv/dr$, $d\rho/dr$, and $dP/dr$ and substitute them above so that Equation \ref{eq:dMdr} simplifies to
\begin{equation}
    \frac{d \mathcal{M}}{dr} = - \frac{ \mathcal{M} ( 2 + (\gamma-1)\mathcal{M}^2) ( r \dot{\mu} ( 1+ \mathcal{M}^2) - 4 v \rho)}{4 (\mathcal{M}^2 -1) r v \rho }.
    \label{eq:dMdr_no_Mdothot_sph} 
\end{equation}
Before any mass-loading, at any radius $r$, the mass-flow rate of the spherical adiabatic wind is 
\begin{equation}
    \dot{M}_\mathrm{hot} = 4 \pi r ^2 \rho_\mathrm{hot} v_\mathrm{hot} = \mathrm{constant}. 
\end{equation}
For consistency with the above equation, we denote the Mach number of the adiabatic flow (before mass-loading) as  $\mathcal{M}_\mathrm{hot}$. We substitute these into Equation \ref{eq:dMdr} and take $\gamma =5/3$ so that 
\begin{equation}
    \frac{d\mathcal{M}_\mathrm{hot}}{dr} = - \frac{2 \mathcal{M}_\mathrm{hot} (\mathcal{M}_\mathrm{hot}^2 +3) ((5 \mathcal{M}_\mathrm{hot}^2 + 3) \pi r^3 \dot{\mu} - 3 \dot{M}_\mathrm{hot})}{9 (\mathcal{M}_\mathrm{hot}^2 - 1 ) r \dot{M}_\mathrm{hot}}.
    \label{eq:dMdr_Mdothot_sph}
\end{equation}
We require mass-loading to be instantaneous and to occur uniformly within an infinitesimally thin shell of volume $V_\mathrm{shell,load} = 4 \pi r^2 \Delta r$ such that the total entrained mass is given by the product of $V_\mathrm{shell,load} \times \dot{\mu}$. The total decrease in Mach number from $r \rightarrow r+\Delta r$ is then
\begin{equation}
    \frac{\Delta \mathcal{M}_\mathrm{hot}}{\Delta r} = - \frac{(\mathcal{M}_\mathrm{hot}-1)}{\Delta r}. 
    \label{eq:dMdr_enforced} 
\end{equation}
We set Equation \ref{eq:dMdr_Mdothot_sph} equal to Equation \ref{eq:dMdr_enforced} and assume the flow is highly supersonic before mass-loading. We find the minimum mass-loading required to decelerate the flow through the sonic point to be 
\begin{equation}
    \dot{M}_\mathrm{load,min} = \frac{18}{5} \frac{\dot{M}_\mathrm{hot}}{\mathcal{M}_\mathrm{hot}^2} \quad (\mathcal{M}_\mathrm{hot} \gg 1).
    \label{eq:Mload_min_sph}
\end{equation}
Equation \ref{eq:Mload_min_sph} shows that the minimum mass-loading criterion decreases with Mach number. We note that this is the Mach number of the supersonic flow before mass-loading. The relevant radial scalings for an adiabatic supersonic spherical wind are $\rho_\mathrm{hot} \propto r^{-2}$, $v_\mathrm{hot} \propto r^0$, and $\mathcal{M}_\mathrm{hot} \propto r^{2/3}$. This means $\mathcal{M}_\mathrm{hot} \propto \rho_\mathrm{hot}^{-1/3}$. Restated again for clarity: higher Mach number flows for a time-steady spherical \textit{adiabatic} wind do not imply higher bulk velocities, it merely means the flow is cooler and less dense. From momentum conservation, for fixed velocity,  we would expect denser winds to require more mass-loading for equivalent deceleration. From Equation \ref{eq:Mload_min_sph}, this appears to be the case as $\dot{M}_\mathrm{load,min} \propto \rho_\mathrm{hot}^{2/3}$.

Uniform feedback launches a wind with a sonic point at the edge of the starburst volume \citep{Chevalier1985}. After leaving this boundary, the flow continues to accelerate due to the steep pressure gradient up until approximately $\mathcal{M} \sim 2$ (e.g., see Fig.~4 of \citet{Nguyen2023}). For an adiabatic spherical wind with $\mathcal{M}_\mathrm{hot} = 2 $, 
\begin{equation}
    \dot{M}_\mathrm{load,min} \sim  \dot{M}_\mathrm{hot} \quad (\mathcal{M}_\mathrm{hot} = 2)
\end{equation}
which is approximately the injected mass within the wind-driving region (see Eq.~\ref{eq:Mdot}).

\section{Simulations}
\label{sec:Chollas} 
In this section we describe our numerical experiments exploring the time-dependent stability of mass-loaded winds. In Section \ref{sec:stable_flows}, we first consider the stability of supersonic spherical mass-loaded flow solutions obtained from the steady-state equations by \citet{Nguyen2021}. These have relatively low mass-loading in the supersonic region and we find them to be stable, as expected from the perturbation analysis. In Section \ref{sec:unstable_flows}, we simulate more heavily mass-loaded flows that are sufficiently decelerated to pass through a sonic point. These solutions become radiative and undergo an instability that produces dense filaments, as sketched in Figure~\ref{fig:doodle}. 

In all of the simulations presented, winds are launched via SN feedback as described by \citet{Chevalier1985} (CC85). Within the starburst volume, $R$, energy and mass are uniformly deposited with rates 
\begin{equation}
    \dot{E}_\mathrm{hot} = 3.1 \times \alpha \times  10^{41} \times \dot{M}_\mathrm{SFR,*} \, \mathrm{[ergs\,s^{-1}]}, 
    \label{eq:Edot}
\end{equation}
and 
\begin{equation}
    \dot{M}_\mathrm{hot} = \beta \times \dot{M}_\mathrm{SFR},  
    \label{eq:Mdot}
\end{equation}
where $\dot{M}_\mathrm{SFR,*} = \dot{M}_\mathrm{SFR}/\mathrm{M_\odot \, yr ^{-1}}$, $\dot{M}_\mathrm{SFR}\,[M_\odot \, yr^{-1}]$ is the star-formation rate, and $\alpha$ and $\beta$ are the dimensionless SNe energy and mass-loading efficiencies, respectively, and we have assumed that there is one supernova per 100 $\mathrm{M_\odot}$ of star formation and that each supernova releases $10^{51}\,$ergs. Beyond the injection radius $R$, the flow is supersonic \citep{Chevalier1985}. At a specified distance away from the wind-driving region $R_\mathrm{load} > R$, we mass-load the flow by sub-grid injection of mass, meant to represent the entrainment of cool low column density clouds into the hot supersonic flow \citep{Cowie1981,Nguyen2021}. We use a single power-law to parameterize the mass-loading:  
\begin{equation}
    \dot{\mu} = \dot{\mu}_0 \times (\lambda / r)^\Delta \quad (R_\mathrm{load} > R), 
    \label{eq:mudot}
\end{equation}
where $\lambda\,\mathrm{[kpc]}$ is an arbitrary length scale, $\Delta$ is the slope of the mass-loading power-law, and $\dot{\mu}_0 \, \mathrm{[M_\odot \, kpc^{-3} \, yr^{-1}]}$ is a constant. The power-law mass-loading is physically motivated by the \citet{Nguyen2021} fit to the CGOLS IV simulation \citep{Schneider2020}. Note that the reader should not confuse mass-loading of the type we explore here (Eq.~\ref{eq:mudot}), with the mass-loading that occurs within the central driving region ($r<R$) that produces the hot outflow (Eq.~\ref{eq:Mdot}). We adopt the cooling curve given by \citet{Schneider2018}, which is an analytic fit to a solar-metallicity collisional ionization equilibrium curve calculated with Cloudy \citep{Ferland2013}. This cooling curve assumes a temperature floor of $10^4$\,K to account for heating and photoionization from UV sources. Throughout the paper we take $\mu = 0.6$ for a plasma with solar metallicity.

\subsection{Stable Mass-loaded Supersonic Flows}
\label{sec:stable_flows}
\begin{figure}
    \centering
    \includegraphics[width=\columnwidth]{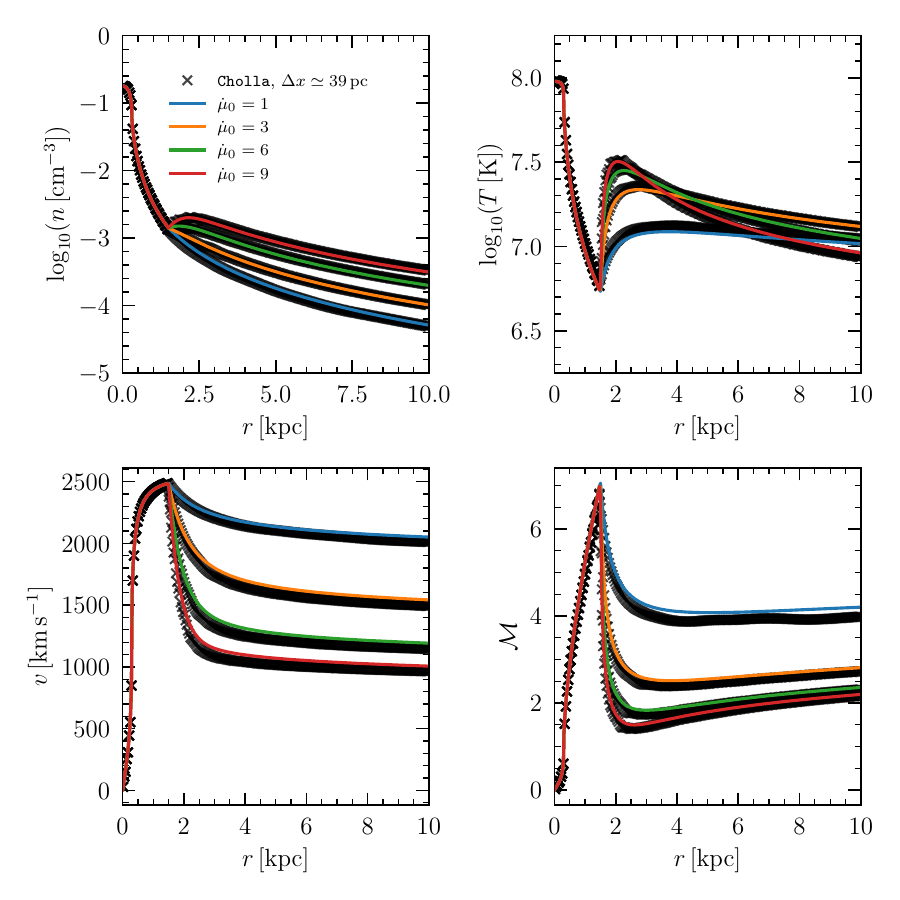}
    \caption{Number density, temperature, velocity, and Mach number profiles for the 1D steady-state semi-analytic wind models (colored lines, see \citet{Nguyen2021}) and $+\hat{z}$ skewers of the time-dependent 3D Cholla simulations (black x's) for mass-loaded spherical wind models with varied volumetric mass-loading rates $\dot{\mu}_0 = 1, \ 3, \ 6, \ \mathrm{and} \ 9 \, \mathrm{M_\odot \, yr^{-1} \, kpc^{-3}}$. These correspond to total mass-loading rates $\dot{M}_\mathrm{load} \simeq  0.5, \ 1.5, \ 3.0, \ \mathrm{and} \ 4.5 \,\mathrm{M_\odot \, yr^{-1}}$, respectively. The \texttt{Cholla} simulations have a cell-width of $\Delta x \simeq 39\,$pc and all quickly reach steady-state and agree well with the 1D models, confirming the time-steady assumption.}
    \label{fig:hotwinds_stable}
\end{figure}
\begin{figure}
    \centering
    \includegraphics[width=\columnwidth]{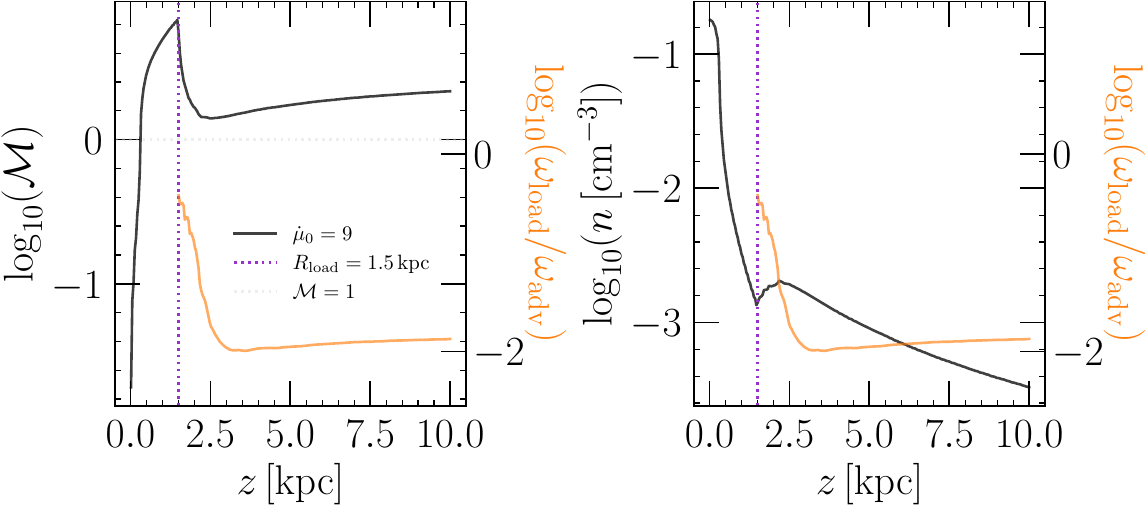}
    \caption{The Mach number, density, and ratio between the instability growth rate (Eq.~\ref{eq:spherical_growthrate}) and the advection rate ($\omega_\mathrm{adv} = v/r$) of the highest mass-loaded supersonic model shown in Figure~\ref{fig:hotwinds_stable}. The dotted purple line is the radius where mass-loading starts. Negative values of $\omega_\mathrm{load}/\omega_\mathrm{adv}$ correspond to instability damping rates. The mass-loading produces small wiggles at the loading radius that are quickly damped out. For this wind model, the minimum Mach number beyond the loading radius is $\mathcal{M} \simeq 1.4$, and the flow does not cross the instability critical points.}
    \label{fig:hotwinds_mudot9_stable}
\end{figure}
We first test the stability of spherical mass-loaded flows that were studied under steady-state assumption in \citet{Nguyen2021}. The simulations span $5\times5\times20\,\mathrm{kpc}$ with $128\times128\times512\,\mathrm{cells}$, giving a cell resolution of $\Delta x \simeq 39\,\mathrm{pc}$. The wind launching parameters (see Eqs.~\ref{eq:Edot} and \ref{eq:Mdot}) are: $\alpha=1, \ \beta=0.15, \ R = 0.3\,\mathrm{kpc}, \ \dot{M}_\mathrm{SFR,*}=10$. We use $\lambda = R$, $R_\mathrm{load} = 1.5\,\mathrm{kpc}$, and $\Delta=3.1$, with varied volumetric mass-loading rates of $\dot{\mu}_0 = 1, \ 3, \ 6, \ \mathrm{and} \ 9 \, \mathrm{M_\odot \, kpc^{-3} \, yr^{-1}}$ in Equation \ref{eq:mudot}. The total mass-loading rates are approximately 0.5, 1.5, 3.0, and 4.5 $\mathrm{\dot{M}_\mathrm{\odot} \, yr^{-1}}$, respectively. 

The simulations all reach a steady state. In Figure \ref{fig:hotwinds_stable}, we plot the time-independent 1D semi-analytic solutions (colored lines) alongside the $+\hat{z}$ skewers of the 3D time-dependent \texttt{Cholla} simulations. We find good agreement between the two, confirming the time-steady assumption of supersonic non-radiative mass-loaded flows studied in previous works \citep{Cowie1981,Suchkov1996,Nguyen2021,Fielding2022}. 

In Figure \ref{fig:hotwinds_mudot9_stable}, we plot 1D Mach number and density skewers (black lines) of the most heavily mass-loaded model from Figure \ref{fig:hotwinds_stable} with   $\dot{\mu}_0=9\mathrm{M_\odot\,yr^{-1}\,kpc^{-3}}$. We also plot the ratio between the instability growth rate (Eq.~\ref{eq:planar_growthrate}) and the advection rate ($\omega_\mathrm{adv}=v/r$) in orange. For $\omega_\mathrm{load}/\omega_\mathrm{adv} = 1$, we expect the instabilities to develop. For $\omega_\mathrm{load}/\omega_\mathrm{adv} < 0$, the instability growth rate is exponentially damped. We see that this ratio is always negative. Strong damping will lead to stable flows (colored lines, Fig.~\ref{fig:hotwinds_stable}). 

The total mass injected into the flow, $M_\mathrm{load} = \int_{R_\mathrm{load}}^{10\,\mathrm{kpc}} dr \, 4 \pi r^2 \dot{\mu}$, in each model exceeds the derived mass-loading criterion (Eq.~\ref{eq:Mload_min_sph}), yet the flow does not transition the sonic point. This is not surprising, because the condition in Equation \ref{eq:Mload_min_sph} is derived under the assumption that mass-loading occurs in an infinitely thin shell. If we, instead, consider the mass-loading between a thin shell bounded by $R_\mathrm{load}$ and $R_\mathrm{load} + \Delta r $, where $\Delta r = \Delta x = 39\,\mathrm{pc}$ (i.e., the cell resolution), then we see that even the strongest mass-loading model (Fig.~\ref{fig:hotwinds_stable} red line, $\dot{\mu} = 9\, \mathrm{M_\odot \, yr^{-1} \, kpc^{-3}}$) does not satisfy the criterion, and thus is consistent with the condition given by Equation \ref{eq:Mload_min_sph}.

\subsection{Unstable Spherical Flows}
\label{sec:unstable_flows}

\begin{figure*}
    \centering
    \includegraphics[width=\textwidth]{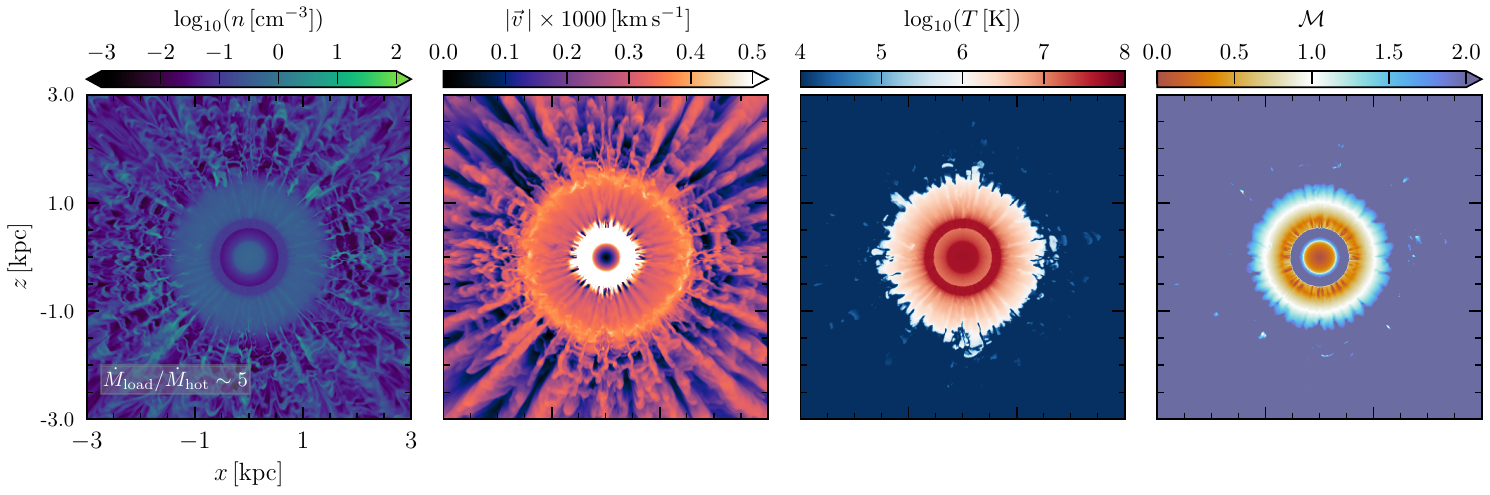}
    \caption{2D density, velocity, temperature, and Mach number slices through $y=0$ for the most heavily mass-loaded model ($\dot{M}_\mathrm{load} / \dot{M}_\mathrm{hot} \sim 5 )$. The supersonic spherical wind is driven from energy and mass injection within a radius $R=0.3\,$kpc. At a radius $R_\mathrm{load}=0.7\,$kpc, the flow is injected with mass (see Eq.~\ref{eq:mudot}). This introduces a stationary rarefaction shock. Mass-loading leads to an unstable flow, where the excited perturbations (see the cornea-like structure in the right two plots) near the loading radius $R_\mathrm{load}$ are quickly damped. These damped perturbations grow non-linearly when the flow becomes radiative and rapidly cools down to $10^4\,$K, leading to the formation of dense filaments as previously seen in the planar case.}
    \label{fig:spherical_2D}
\end{figure*}
\begin{figure*}
    \centering
    \includegraphics[width=\textwidth]{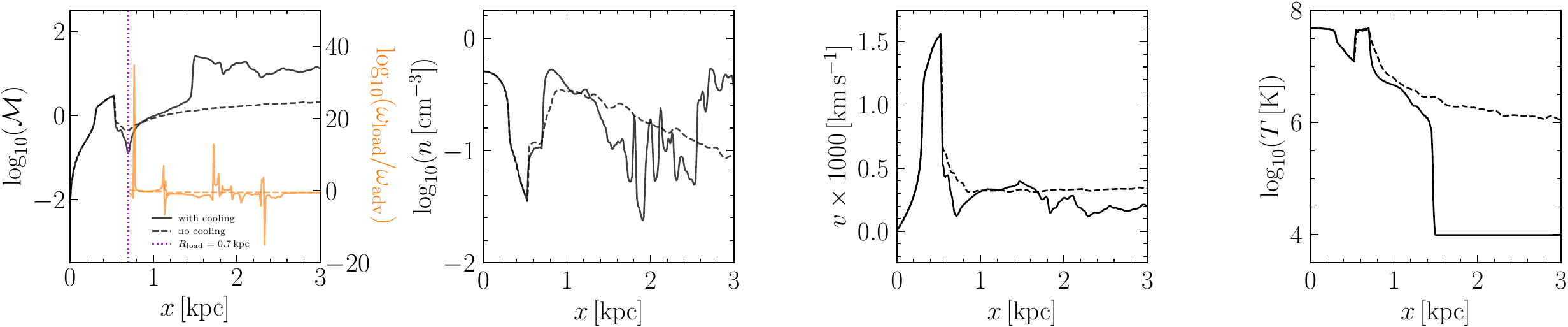}
    \caption{The Mach number, density, velocity, temperature, and ratio between the instability growth rate (Eq.~\ref{eq:spherical_growthrate}) and the advection rate ($\omega_\mathrm{adv} = v/r$) for the heavily mass-loaded spherical flow with and without radiative cooling. We find mass-loading in both the cooling and no-cooling cases results in fluctuations in the density, however these fluctuations can seed a larger scale thermal instability in the cooling case, leading to the filamentary structures apparent in the 2D slices.}
    \label{fig:MnTv_unstable}
\end{figure*}

 We now study the behavior of heavily mass-loaded wind models that reach a sonic point. This regime of large mass-loading was not explored in our earlier work \citep{Nguyen2021}. The simulations span $6\times6\times6\,\mathrm{kpc}$. The fiducial models are run at $512^3$ cells, giving a cell resolution of $\Delta x \simeq 11.7\,\mathrm{pc}$. The wind launching parameters are: $\alpha = 1.0, \ \beta =0.3, \ R = 0.3\,\mathrm{kpc}\,$ and $\dot{M}_\mathrm{SFR,*}=10$. The mass-loading parameters are (see Eq.~\ref{eq:mudot}) $\lambda = R_\mathrm{load} = 0.7\,\mathrm{kpc}$, $\Delta=4$, and $\dot{\mu}_0 = 2, \ 6,  \ 10, \ \mathrm{and} \ 15\,\mathrm{[M_\odot \, kpc^{-3} \, yr^{-1}]}$. We do not introduce any perturbations into the simulations by hand. However, there are grid effects when simulating a spherical wind on a Cartesian grid that give rise to perturbations in the flow.

In Figure \ref{fig:spherical_2D} we plot snapshots of density, temperature, velocity, and Mach number for the most heavily mass-loaded wind, after the flow is in a statistical steady-state. We find a stationary shockwave and radial high-density filaments. Unlike the stable mass-loaded winds discussed in the previous section, the winds here cross the sonic point, as shown in the Mach number panel, where the green-blue color indicate the flow is substantially subsonic ($\mathcal{M} \ll 1 $). Before the mass-loading radius $R_\mathrm{load}=0.7$\,kpc, there is a clear shockwave that is time-steady (for constant $\dot{\mu}$, $\dot{E}_\mathrm{hot}$, and $\dot{M}_\mathrm{hot}$. After being mass-loaded, the flow undergoes rapid bulk cooling down to $10^4\,$K and forms dense filaments. 

\begin{figure*}
    \centering
    \includegraphics[width=\textwidth]{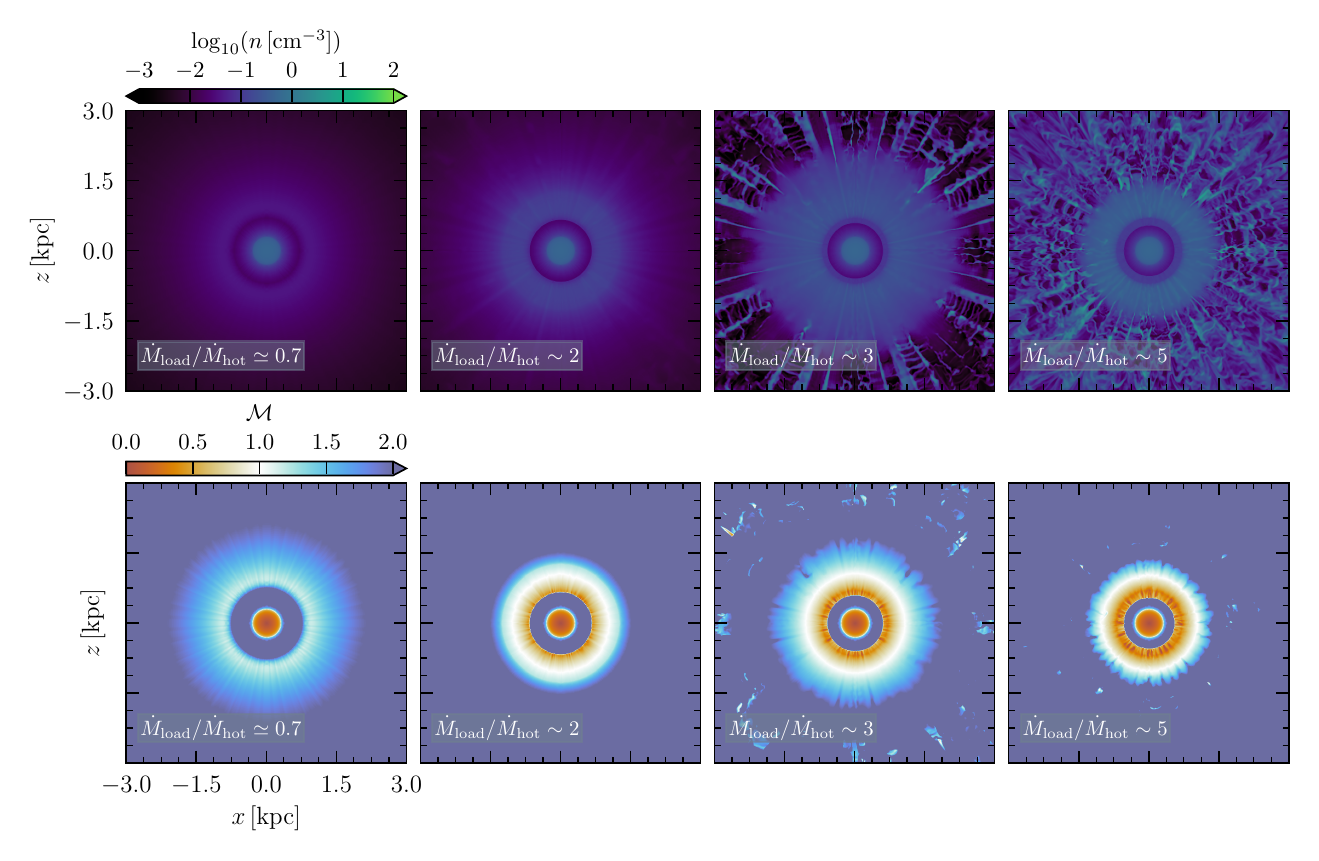}
    \caption{2D density slices for varied mass-loading rates $\dot{\mu}_0=2,\ 6, \, 10, \ \mathrm{and} \ 15 \, \mathrm{[M_\odot \, yr^{-1} \, kpc^{-3}]}$ which correspond to mass-loading ratios of $\dot{M}_\mathrm{load}/\dot{M}_\mathrm{hot} \sim 0.7, \, 2, \, 3, \ \mathrm{and} \ 5 $, respectively. We see that the three most mass-loaded models develop sonic points (orange rings), as expected from the derived criterion of  $\dot{M}_\mathrm{load}/\dot{M}_\mathrm{hot} \gtrsim 1$ (Eq. \ref{eq:Mach_crit}). Higher mass-loading increases the gas density, causing it to become radiative on smaller scales. Filaments form when the gas undergoes rapid bulk cooling, which is apparent in the $\dot{M}_\mathrm{load}/\dot{M}_\mathrm{hot} \sim 3 \ \mathrm{and} \ 5$ simulations. The cooling radius is larger than the simulation box for the $\dot{M}_\mathrm{load}/\dot{M}_\mathrm{hot} \sim 2$ model, but we can see small fluctuations in the density profile that are expected to develop into larger-scale thermal instability after rapidly cooling. These small fluctuations are not present in the lowest mass-loaded model which remains stable, supersonic, and non-radiative.}
    \label{fig:512_all_mudots}
\end{figure*}

What is the origin of the observed non-linear wind evolution? Single phase, radiatively-cooling outflows have been shown to be stable \citep{Schneider2018}, and do not lead to the dense filament formation that we observe here. During the deceleration through $\mathcal{M} \sim 1$, the flow develops density perturbations because of the analysis shown in \ref{sec:linear_stability}. However, these do not become non-linear without radiative cooling. 

We plot the 1D $\hat{x}_+$ profiles in Figure \ref{fig:MnTv_unstable}. The purple vertical line shows the loading radius $R_\mathrm{load}$. Every physical quantity shows a jump well before the mass-loading loading radius. At the start of the simulation, a shockwave forms at $R_{\rm load}$, but then settles at a radius smaller than $R_\mathrm{load}$ because of the thermal pressure of the subsonic material.  Mass-loading for this particular skewer drives the flow to subsonic speeds of $\log_{10}(\mathcal{M}) \sim 0.5$, but from the 2D slices shown in Figure \ref{fig:spherical_2D}, we see that the flow decelerates even to $\log_{10}(\mathcal{M}) \sim -2$ along some radial directions. In Figure \ref{fig:MnTv_unstable}, the orange line shows the ratio of the competing rates, where $\omega_\mathrm{load}/\omega_\mathrm{adv} \gg 1$ implies instability growth. This ratio peaks precisely at the sonic point, as expected from Figure \ref{fig:growth_rates}. These perturbations are damped, but when combined with radiative cooling are unstable to filament formation via thermal instability. This is explicitly shown in the dashed line, which is an identical simulation, but, without radiative cooling turned on. There are small fluctuations in each quantity, but for the most part, the flow returns to a stable configuration. It is only when we turn on radiative cooling, that these fluctuations grow and thermal instability takes over. We attribute the small fluctuations to mass-loading through the sonic point (see Sec.~\ref{sec:linear_stability}). These fluctuations are not present in the simulated models shown in the previous section, which are stable and supersonic at all radii. 

In Figure \ref{fig:512_all_mudots}, we plot 2D slices for density and Mach number for the four different values of mass-loading $\dot{\mu}_0=2,\ 6, \, 10, \ \mathrm{and} \ 15 \, \mathrm{[M_\odot \, yr^{-1} \, kpc^{-3}]}$ which correspond to mass-loading ratios of $\dot{M}_\mathrm{load}/\dot{M}_\mathrm{hot} \sim 0.7, \, 2, \, 3, \ \mathrm{and} \ 5 $, respectively. We see that for the $\dot{M}_\mathrm{load}/\dot{M}_\mathrm{hot} \simeq 0.7$ simulation (first column), the flow does not cross a sonic point and remains non-radiative and stable. This is explicitly seen in the Mach number panel. For higher values of the mass-loading parameter, the Mach number decreases into the subsonic regime (identified by the orange ring in each panel). The cooling radius is set by the total mass swept up, and for higher mass-loading parameters the flow rapidly cools and forms dense filaments within the simulation box. For $\dot{M}_\mathrm{load}/\dot{M}_\mathrm{hot} \simeq 2$, the cooling radius exceeds the simulation domain, but we can still see small variations in the density that are expected to seed the larger-scale thermal instability. These small fluctuations are not seen in the lowest mass-loaded model because it does not cross the sonic point.

\begin{figure*}
    \centering
    \includegraphics[width=\textwidth]{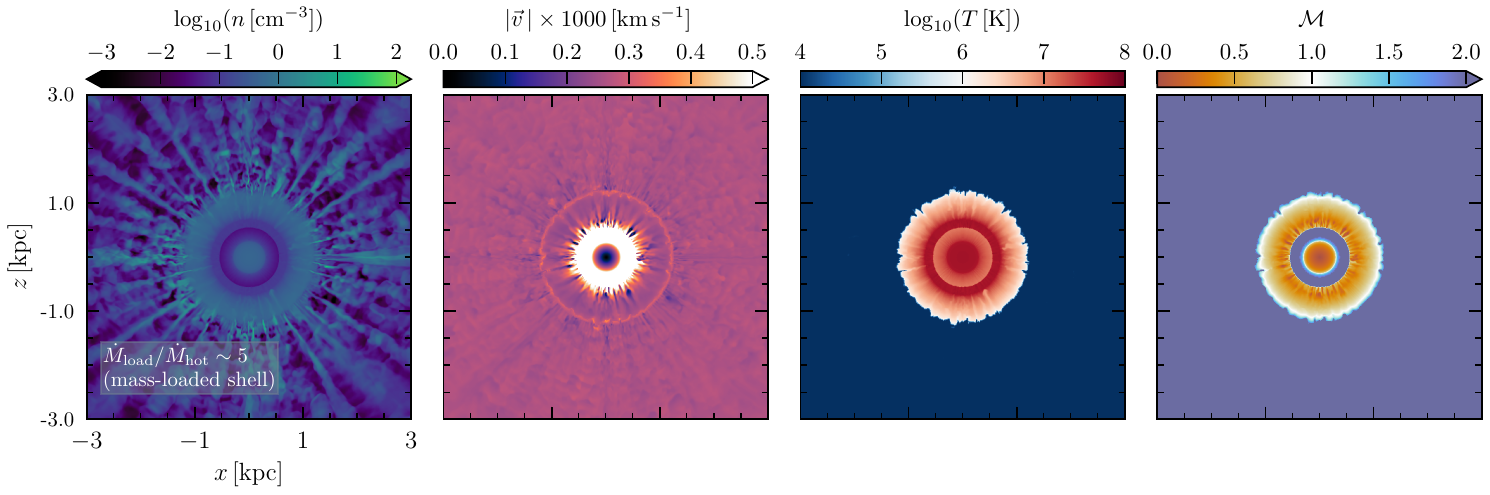}
    \caption{2D density, velocity, temperature, and Mach number slices through $y=0$. The wind is driven from energy and mass injection within a radius $R=0.3\,$kpc. Between radii $R_\mathrm{load}=0.7 \leq r \leq 1.25\,$kpc the flow is uniformly injected with total mass equivalent to that injected for the simulation presented in Figure \ref{fig:spherical_2D}. Similar to the distributed mass-loading scenario, filaments form after rapidly cooling, however the velocity structure is nearly constant, with filaments moving slightly slower than voids.} 
    \label{fig:mudot_uni_1x4}
\end{figure*}

\begin{figure}
    \centering
    \includegraphics[width=\columnwidth]{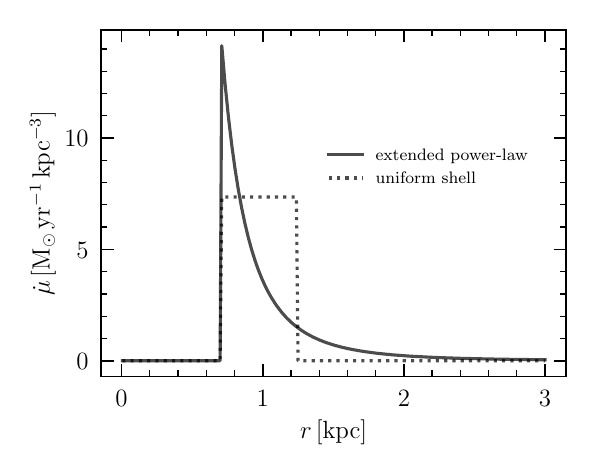}
    \caption{The volumetric mass-loading rates for an extended powerlaw ($\dot{\mu} \propto r^{-\Delta}$ for $0.7\,\mathrm{kpc} \leq r$) and mass-loading with sharp drop off ($\dot{\mu} = \mathrm{const}$ for $0.7 \leq r \leq 1.25\,$kpc). Both of these mass-loading functions lead to filament formation (see Figs.\ref{fig:spherical_2D} and \ref{fig:mudot_uni_1x4}), however the the filaments have different kinematics (see Fig.\ref{fig:hexbins}).}
    \label{fig:mudot_compare}
\end{figure}

\begin{figure}
    \centering
    \includegraphics[width=\columnwidth]{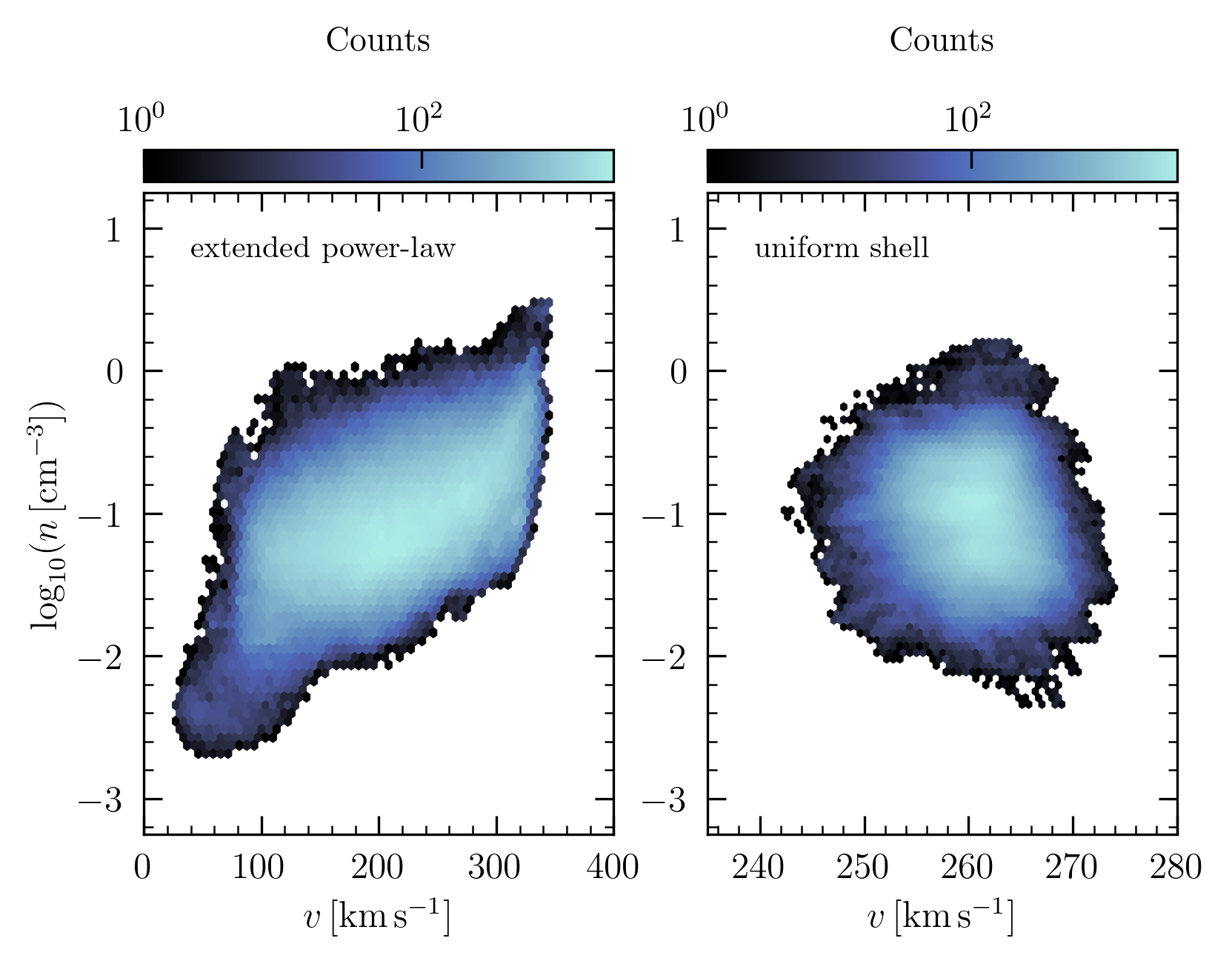}
    \caption{Density vs. velocity distributions for material within a thin 100\,pc shell located at $r_\mathrm{shell} = 2.75\,$kpc for two different mass-loading functions  with equivalent total mass injection: an extended power-law (left) and uniform with sharp drop-off (right), see Figure~\ref{fig:mudot_compare}. For the power-law scenario, we see that the densest material moves the fastest and that the total range of velocities spans several hundred km\,s$^{-1}$ (see Fig.~\ref{fig:spherical_2D}). In contrast, the uniform shell mass-loading function leads to filaments with roughly constant velocity. The highest density components do not move the fastest (see Fig.~\ref{fig:mudot_uni_1x4}). Both power-law mass-loaded and uniform-shell mass-loaded material contain the same average total energy density of $\langle E \rangle = \langle \rho  v  ^2 /2 +  P \rangle \simeq 4 \times 10^{-11} \ \mathrm{ergs \, cm^{-3}}$. }
    \label{fig:hexbins}
\end{figure}

As seen in panels left two panels of Figure \ref{fig:spherical_2D}, the filaments move faster than the surrounding post-radiative cool wind. To investigate the origin of this, we additionally consider uniform mass-loading with sharp drop off. The two mass-loading functions are shown for comparison in Figure \ref{fig:mudot_compare}. In Figure \ref{fig:mudot_uni_1x4}, we plot 2D slices of density, temperature, velocity, and Mach number analogous to the earlier Figure \ref{fig:spherical_2D}. Similar to the extended power-law mass-loading, dense filaments also form. However, the rich velocity structure seen in panel 3 of Figure~\ref{fig:spherical_2D} is not present here (panel 3 of Fig.~\ref{fig:mudot_uni_1x4}). In contrast to the fast filaments seen in the extended power-law mass-loading simulation (Fig.~\ref{fig:spherical_2D}), the velocity is roughly constant, with the dense filaments moving slightly slower. This difference is most explicitly present in the density vs. velocity distributions for each simulation in Figure \ref{fig:hexbins}. Each panel showcases the density vs. velocity by cell count for the two different mass-loading functions. The key take away is that densest filaments move fastest when the mass-loading function is an extended power-law scenario (left panel), whereas they do not when mass-loading is uniform (right panel). Both power-law mass-loaded and uniform-shell mass-loaded material contain the same average total energy density of $\langle E \rangle = \langle \rho  v  ^2 /2 +  P \rangle \simeq 4 \times 10^{-11} \ \mathrm{ergs \, cm^{-3}}$. Despite there being equivalent total mass being ``swept-up", we show that the kinematic structure of the wind is sensitive to how mass-loading occurs.    

\begin{figure}
    \centering
    \includegraphics[width=\columnwidth]{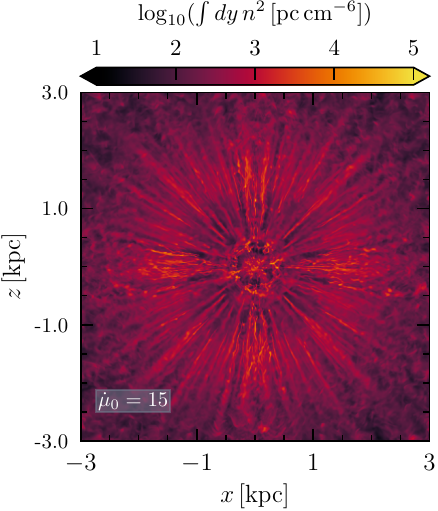}
    \caption{The emission measure of material less than $2 \times 10^4\,K$.}
    \label{fig:nsquared_y_integrated}
\end{figure}

\begin{figure}
    \centering
    \includegraphics[width=\columnwidth]{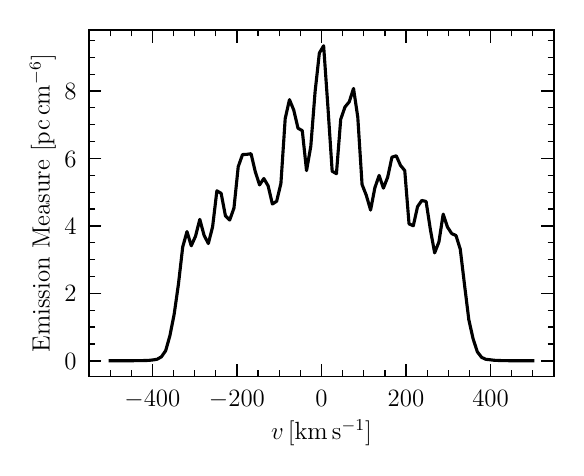}
    \caption{Emission measure as a function of binned velocity for cool, $T \leq 2 \times 10^4\,$K material. There is non-neglible emission of cool filaments moving at velocities of several hundred km\,s$^{-1}$.}
    \label{fig:em_velocity}
\end{figure}

In Figure \ref{fig:nsquared_y_integrated}, we plot the projected emission measure of cool material for the most heavily mass-loaded extended power-law model ($\dot{\mu}_0 = 15\,\mathrm{M_\odot \, yr^{-1} \, kpc^{-3}}$). We exclude all material that has $T \geq 2 \times 10^4\,$K, which is essentially all of the hot material within the cooling radius shown in the temperature panel of Figure \ref{fig:spherical_2D}. The projected H$\alpha$ intensity can then be obtained by multiplying the material's emission measure by the appropriate recombination  rate (see Eq.~14.8 of \citealt{Draine2011}), which we do not calculate here. The emission measure of the filaments are pronounced and reminiscent of the H$\alpha$ emitting structures seen in M82 \citep{Lehnert1999,Newman2012}. In Figure \ref{fig:em_velocity}, we plot the emission measure as a function of velocity for the same cool material, integrated over the computational domain. We see that the filaments contribute to the emission measure and would be apparent in resolved velocity line measurements at several hundred km\,s$^{-1}$ (e.g., \citealt{Shopbell1998,Shapiro2009}). A more quantitative discussion of the filament dynamics, projected column density distributions, and comparison with observations is saved for a future work. More discussion is provided in Section \autoref{sec:discussion}).

\begin{figure*}
    \centering \includegraphics[width=0.9\textwidth]{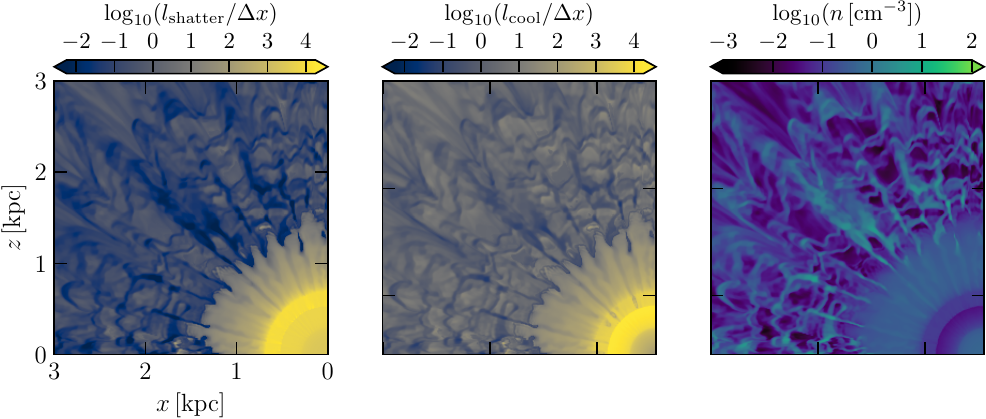}
    \caption{2D quarter wedge slices for the shattering length, cooling length, and density for extended power-law mass-loading simulation (Fig.~\ref{fig:spherical_2D}). Here the spatial resolution is $512^3 \mathrm{ \ cells}  \rightarrow \Delta x \sim 11.7 \, \mathrm{pc}$. Both the shattering length and cooling length are unresolved in the dense filaments by at least two orders of magnitude.}
    \label{fig:characteristic_lengths}
\end{figure*}

\begin{figure*}
    \centering
    \includegraphics[width=\textwidth]{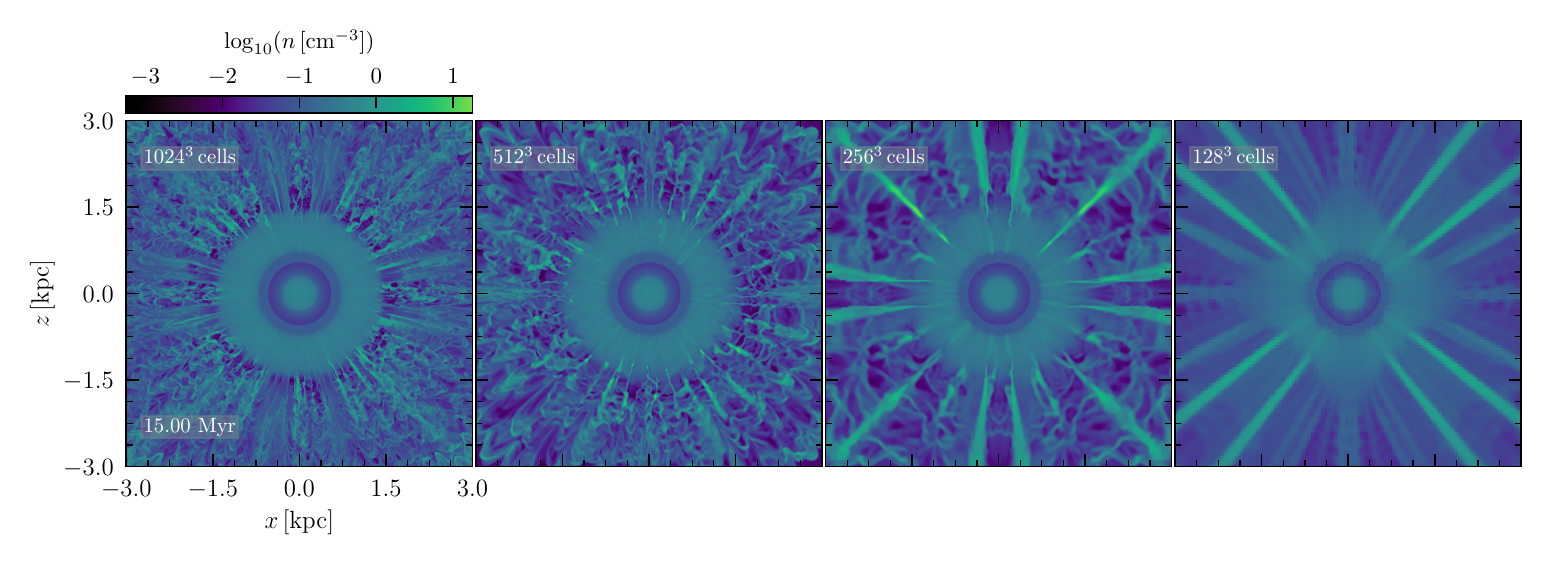}
    \caption{2$d$ density slices for four different resolutions of $1024^3, \ 512^3, \ 256^3, \ \mathrm{and} \ 128^3$ cells. Because the simulation is run on a Cartesian grid, the numerical artifacts are stronger in the lower resolutions when simulating a spherical system. More prominent grid effects lead to stronger perturbations. These are localized as seen in the ``X" shape, where the filaments are formed (see right two panels). As the resolution increases, the grid effects are decreases, and we see a more isotropic spatial distribution of filamentary formation. }    
    \label{fig:resolution}
\end{figure*}

We have tested the resolution dependence of the qualitative features of the unstable mass-loaded models. In Figure \ref{fig:resolution}, we plot four different resolutions ($1024^3$, $512^3$, $256^3$, and $128^3$ cells) of the heavily mass-loaded power-law spherical wind simulation (Fig.~\ref{fig:spherical_2D}, $512^3$ resolution). Because \texttt{Cholla} uses a Cartesian grid, there are numerical grid effects when simulating a spherical wind, with larger effects at lower resolutions. The higher the resolution, the more isotropic the filament formation, and the filaments appear smaller. Although the perturbations in these simulations are seeded by numerical effects, in the realistic environments one would expect similar, if not stronger, effects simply from the non-uniform nature of feedback \citep{Fielding2017,Tan2023b}, turbulence within the ISM \citep{Lynn2012}, fractal inter-cloud structure of the ISM \citep{Elmegreen1997,Tanner2016}, among other possibilities. As shown in Figure \ref{fig:characteristic_lengths}, we do not resolve the shattering scale $l_\mathrm{shatter}$ \citep{McCourt2018}:
\begin{equation}
    l_\mathrm{shatter} = \mathrm{min}(c_s\,t_\mathrm{cool}).
    \label{eq:shatter}
\end{equation}
nor the cooling scale
\begin{equation}
    l_\mathrm{cool} = \mathrm{min}(v\,t_\mathrm{cool}).
    \label{eq:cool}
\end{equation}
As shown in Figure \ref{fig:characteristic_lengths}, the shattering and cooling lengths are smallest at the base of the dense filaments, reaching values as small as $l_\mathrm{shatter} \simeq 0.0005\,$pc and $l_\mathrm{cool} \simeq 0.09\,$pc, factors of $\sim 2500$ and $\sim 100$ smaller than the grid scale. Both scales are much larger in the cool medium between the filaments, with typical values relative to the grid scale of $l_\mathrm{shatter} \simeq 370\,$kpc and $l_\mathrm{cool} \simeq 290\,$kpc, respectively. Our highest resolution simulation ($1024^3, \ \Delta x = 5.85$ pc) misses shattering scale by several orders of magnitude. 

\section{Discussion}
\label{sec:discussion}

\subsection{Observational implications}
\citet{Thompson2016} postulated that the rapid destruction of cool clouds near the wind-driving region would lead to mass-loading of the hot phase and then a radiative instability that would lead to the reformation of cool wind material at a larger scale, in a process they termed ``cool cloud transmigration." Here, we show that the mass-loading of cool gas into the hot flow combined with radiative cooling produces an instability identified in Section \ref{sec:unstable_flows}. By forcing the flow through a sonic point, we show that mass-loading provides a mechanism for the cool gas to spontaneously form dense filaments and produce distributions of velocity, density, and column density (Fig.~\ref{fig:hexbins}) that differ markedly from the expectations of a spherical adiabatic outflow or from models that accelerate cool clouds via ram pressure or other mechanisms. This model produces dense filaments of cool material that should produce strong recombination emission (Fig.~\ref{fig:nsquared_y_integrated}), possibly explaining the prominent fast H$\alpha$-emitting filaments in the galactic outflows of ultra-luminous infrared galaxies \citep{Soto2012,Martin2015}, nearby starbursts such as M82 \citep{Shopbell1998,Lehnert1999,Strickland2000,Strickland2004a} and NGC 253 \citep{Arnaboldi1995,Rodriguez-Rico2006}, and high-redshift star-forming galaxies \citep{Shapiro2009,Newman2012}. 

As we show in Figure \ref{fig:nsquared_y_integrated}, when we project the emission along the line of sight, the dense filaments are prominent, with kinematics of order several hundred km\,s (see Figs.~\ref{fig:MnTv_unstable} and \ref{fig:em_velocity}). These velocities are qualitatively similar to the kinematics of M82's H$\alpha$ filaments ($v  \sim 500 - 650 \, \mathrm{km \, s^{-1}}$) observed by \citet{Shopbell1998}, who additionally find a roughly constant velocity gradient. The 1D skewers of the filaments (Fig.~\ref{fig:appendix}) indicate that the filaments  are also moving at roughly constant velocity for the power-law mass-loading model (see Fig.~\ref{fig:appendix}, Section \ref{sec:appendix}). Although we have carried out only a small parameter study, we find that the kinematics of the filaments depends on the spatial distribution of the mass-loading (see Fig.~\ref{fig:hexbins}). In other wind models, predictions for the velocity gradient $dv/dr$ for the cool material are determined by other physics: e.g., ram pressure acceleration of the hot gas on the cool clouds \citep{Zhang2017,Fielding2020}, momentum and energy exchange from between the hot and cool fluids \citep{Gronke2020,Schneider2020,Fielding2022}, cosmic ray acceleration \citep{Ipavich1975,Farber2018, Quataert2022a,Quataert2022b}, and/or radiation pressure \citep{Thompson2015,Blackstone2023,Menon2023}. As we show in Figure \ref{fig:hexbins}, despite equivalent total mass-deposition, the spatial dependency of the mass-loading function impacts filament kinematics in a non-linear manner that requires future investigation.

\subsection{Stability for central vs. distributed mass-loading}
\label{sec:central_vs_distributed}
The fate of a large-scale hot wind is determined by the amount of cool material is entrained. \citet{Wang1995,Silich2004,Thompson2016} showed that if the mass-loading in the central wind-driving region is high enough, the flow rapidly cools at $r_\mathrm{cool}$ down to $10^4\,K$. This picture is qualitatively similar to cool cloud entrainment \textit{outside} of the wind-driving region that we consider throughout this work and in \citet{Nguyen2021}, however there are some important differences. First, the central mass-loading considered by \citet{Wang1995,  Suchkov1996, Silich2004, Thompson2016} occurs within the wind-driving region, implying the added mass has a significant energy content and that the initial flow being mass-loaded is subsonic. \citet{Smith1996} found that mass-loading with additional energy injection generically leads to a stable rising Mach number solution (like that in CC85). This may be why galactic winds that undergo rapid cooling due to heavy central mass-loading were found to be stable \citep{Schneider2018}. In this work we showed that mass-loading with no energy content can lead to instability, in agreement with the prediction of \citet{Smith1996}.

\subsection{Implications for mixing on smaller scales} 
The process of mass-loading is prevalent in the microphysical interactions within multi-phase media. Recent works show that there are regions of parameter space where cool clouds can grow from a mixing cooling cycle in wind tunnel configurations \citep{Gronke2018,Sparre2019,Sparre2020,Gronke2020}, in turbulent backgrounds from coagulation \citep{Gronke2022}, in tails of ram-pressure stripped galaxies \citep{Tonnesen2021}, during gravitational infall \citep{Tan2023a}, and combinations of all of the above \citep[see][for a review]{Faucher-Giguere2023}. Mixing occurs at the interface between the hot wind and cool cloud, where, if the cooling-efficiency at the turbulent mixing layer is sufficiently large, then mixed `warm' gas can contribute to new co-moving cool gas \citep{Begelman1990,Fielding2020,Tan2021,Abruzzo2022}. Critical to this process is the transfer of momentum from the hot phase to the cool phase, which may accelerate a cloud \citep{Gronke2020,Schneider2020} or decelerate an in-falling cloud \citep{Tan2023a}. In the former case, the transfer of momentum requires the rapid deceleration of hot supersonic material as it mixes with the surface layer of the static cool cloud. In this work, we show that strong deceleration of the hot phase through the sonic point may promote thermal instability and formation into filaments, perhaps being an important piece of physics in understanding how hot gas mixes, cools, and condensates onto clouds on the microphysical scale.  

\subsection{Caveats}
The simulations presented here are idealized configurations meant to explore solely the stability of mass-loaded flows. In doing so, we have neglected physics that is present in realistic galactic environments. In particular, we assume a single cooling function fit to solar composition gas that collisional ionization equilibrium, which may not be true for  extended low density outflows \citep{Gray2019a,Gray2019b,Sarkar2022}. We do not include conduction, which may dominate energy transport in the winds before they are mass-loaded \citep{Thompson2016}. Furthermore, in our treatment of background UV heating, our cooling curve truncates at $T_\mathrm{floor} = 10^4\,\mathrm{K}$, but the scale of cold filament formation and dynamics of hot gas changes if the gas is allowed to cool to $10\,\mathrm{K}$ \citep{Tanner2016}. This effect may have implications for the dynamics and physical size of the filaments we find here. We do not resolve the shattering scale-length $l_\mathrm{shatter}$ \citep{McCourt2018} in any of our simulations (See Fig.~\ref{fig:characteristic_lengths}), implying our simulations do not include any shattering-like behavior which may be important in setting the physical size of filaments. We have assumed that the thermal and kinetic energy of the swept up cool clouds to be zero which may have implications for flow stability \citep{Smith1996}. Additionally, the $R_\mathrm{load}$ we chose is more than twice the starburst radius $R$, however \citet{Schneider2020} found mass-loading to be quite close to the wind-driving region. Moving the loading radius inwards closer to $R$ implies the hot flow will be closer to the sonic point (before mass-loading), which may have implications for filament formation.  Furthermore, the dynamics of the cool flows and filaments shown here may be different if the cooling length is resolved \citep{Rey2023}. Lastly, the kinematics of the filaments are expected to be different in the presence of a  gravitational potential, which we neglect here. 

\section{Synthesis} 
\label{sec:conclusion}
Mass-loading is expected to occur on a wide range of scales from hot galactic superwinds entraining clouds, to the microphysical intricacies of multi-phase mixing. Here, we considered the stability of supersonic hot galactic winds that undergo mass-loading. Our main findings are:  
\begin{enumerate}
    \item The steady-state assumptions of mass-loaded 1D models in previous works \citep{Cowie1981,Nguyen2021,Fielding2022} are valid provided the flows are not decelerated through the sonic point (Fig.~\ref{fig:hotwinds_stable}), in agreement with our linear stability analysis (Sec.~\ref{sec:linear_stability}). 
    \item  Mass-loaded flows are unstable as they pass through critical points $\mathcal{M} = 1$ and $1/\sqrt{3}$ (Fig.~\ref{fig:growth_rates}). When the flow enters the subsonic regime, it is re-accelerated back to supersonic speeds through a saddle-point like transition \citet{Williams1994}. The flow leads to dense filament formation as the flow subsequently undergoes rapid bulk cooling (Fig.~\ref{fig:spherical_2D}). To our knowledge, this instability has not been seen in previous CC85-like wind calculations with radiative cooling, although it may be implicit in simulations like \citet{Cooper2009} and \citet{Schneider2020} where there is a complex interaction between hot and cooler phases. We explicitly show cooling is required for non-linear filament formation (Fig.~\ref{fig:MnTv_unstable}). 
    \item Filaments are expected to produce recombination radiation. If the radiation is optically thin, the filaments projected along the line of sight are structurally coherent, and are qualitatively similar to the H$\alpha$ filaments observed in the galactic winds of starburst M82 (Figs.~\ref{fig:nsquared_y_integrated} and \ref{fig:em_velocity}). 
    \item The kinematics of the filaments depend on the spatial distribution of mass-loading, even if the total mass entrained into the hot wind is equivalent (Fig.~\ref{fig:hexbins}). Abrupt mass-loading leads to filaments that move slower than the surrounding wind. Distributed mass-loading over large scales, leads to filaments that move faster than the surrounding wind. We used a power-law to describe distributed mass-loading, which was also similarly used in fitting mass-loading of the hot wind \citep{Nguyen2021} for the CGOLS IV simulation \citep{Schneider2020}.
    \item The minimum mass-loading required to decelerate a supersonic flow down to the sonic point is of order the injected mass from SNe (Eq.~\ref{eq:Mload_min_sph}), but depends on the spatial extent over which mass-loading occurs. This implies that if hot supersonic flows sweep up enough cool clouds, it would ultimately result in the production of fast cool clouds. We verify this criterion with 3D hydrodynamic simulations (Fig.~\ref{fig:512_all_mudots}). 
\end{enumerate}

\section*{Appendix A: Skewers}
\label{sec:appendix}
\begin{figure*}
    \centering
    \includegraphics[width=\textwidth]{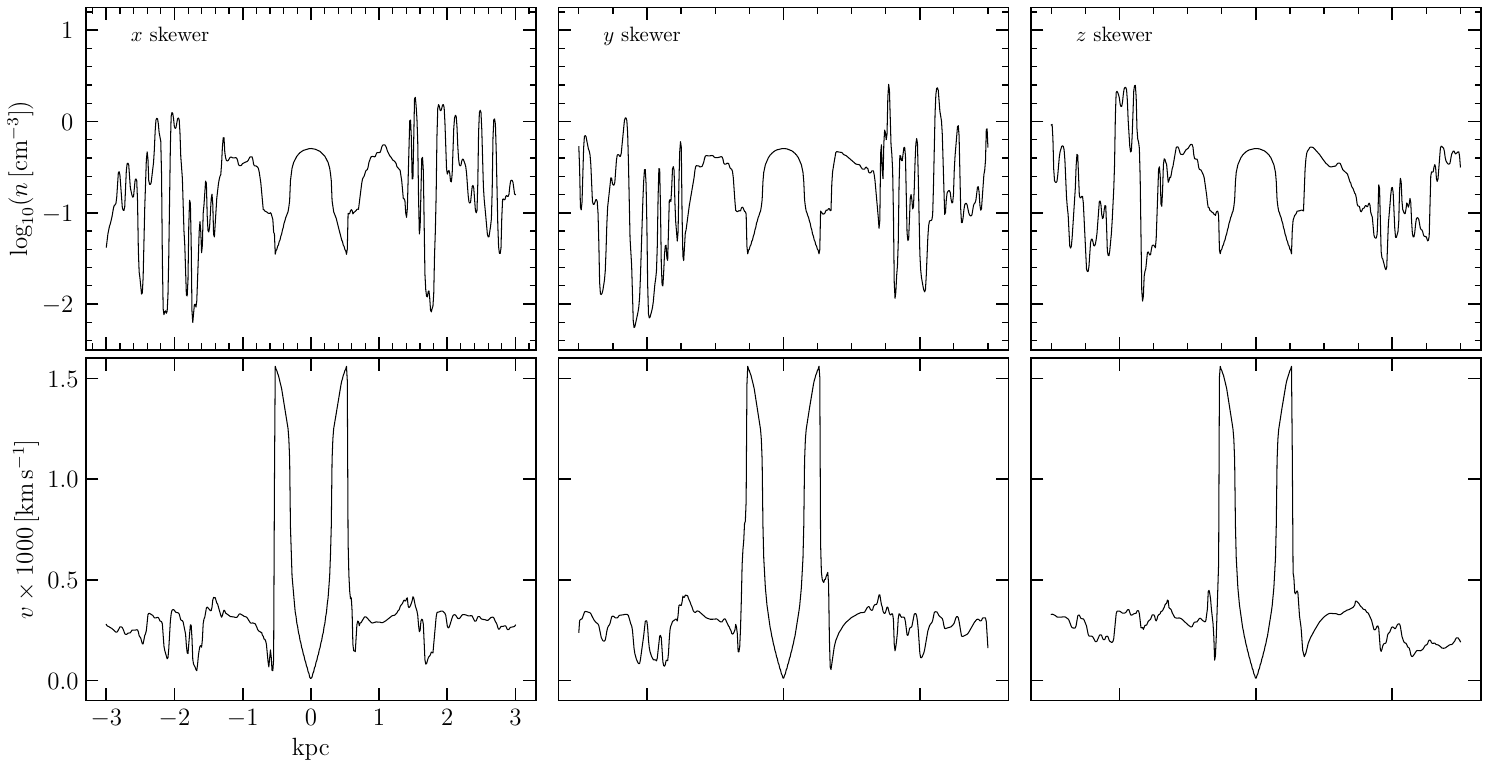}
    \caption{$x$, $y$, and $z$ density and velocity skewers of mass-loaded wind simulation for $\dot{M}_\mathrm{load} / \dot{M}_\mathrm{hot} \sim 5$. The non-linearity of the filaments is explicitly shown in every Cartesian direction, as expected from views of the 2D slices. After cooling to $T\sim 10^4\,$K, the filaments along each skewer exhibit roughly constant velocity of several hundred km\,s$^{-1}$.}
    \label{fig:appendix}
\end{figure*}

In Figure \ref{fig:appendix} we plot density and velocity additional skewers along the $\pm  \ x, \ y, \ \mathrm{and} \ z$ axes to showcase the symmetry of non-linear filament formation which is expected for mass-loading within a shell.  

\section*{Acknowledgements} 
DDN, TAT, and APT thanks the OSU Galaxy Group and Chris Hirata for insightful conversations. DDN and TAT are supported by NSF\,\#1516967, NASA\,ATP\,80NSSC18K0526, and NASA\,21-ASTRO21-0174. E.E.S. acknowledges support from NASA\,TCAN\,80NSSC21K0271 and ATP\,80NSSC22K0720, as well as the David and Lucile Packard Foundation.

\section*{Data Availability}
The data underlying this article will be shared on reasonable request to the corresponding author.



\bibliographystyle{mnras}
\bibliography{bibliography} 



\appendix

\section{Dispersion Relation for Planar Mass-loaded Flow}
\label{sec:planar_disp_relation}
The general dispersion relation for the planar flow is given by 

\begin{align*}
  & \bigg[i ( - i v^6 \gamma (1+ \gamma)^3 \dot{\mu}^3 \rho + 4 k^3 v (P \gamma - v^2 \rho)^4 + 2 v^2(1+\gamma) \dot{\mu}^2(-P \gamma + v^2 \rho) \\
  & (P \gamma^2 - v^2 (1+ \gamma + \gamma^2) \rho) \omega + 2 i \gamma \dot{\mu} (P \gamma -v ^2 \rho)^2 (2P+v^2(3+\gamma) \rho) \omega^2 + \\
  & 4\rho (P\gamma-v^2 \rho)^3 \omega^3 + 2k^2 (P \gamma - v^2 \rho)^2 (i v^2 \gamma (1+\gamma) \dot{\mu}(P+v^2 \rho) - \\
  & 2 (P\gamma - 3 v^2 \rho) (P \gamma - v^2 \rho) \omega ) + k v (- P \gamma + v^2 \rho)(-v^2(1+\gamma) \dot{\mu}^2 (4 P \gamma^2 + v^2 \\ 
  & (3+ \gamma^2) \rho) + 4 i \gamma (2+\gamma) \dot{\mu} (P\gamma - v^2 \rho) (P + v^2 \rho) \omega + 12 \rho (P\gamma - v^2 \rho)^2 \omega^2))\bigg] \\
  & \bigg/ \bigg[2 k(P\gamma -v ^2 \rho)^2 (2 k (P v \gamma - v^3 \rho) - 2 P \gamma \omega  - i v^2 ( \dot{\mu} + \gamma \dot{\mu} + 2 i \rho \omega )) \bigg]
\end{align*}

\section{Comparison to Previous Work}
\label{app:Schekinov}
\citet{Schekinov1996} derived an instability growth rate that scales as $\omega_\mathrm{load} \propto \mathcal{M}^2$ in the high Mach number limit. This disagrees with our derived growth rate, Equation \ref{eq:planar_growthrate}. Taking their general derived growth rate (Eq.~12 of \citet{Schekinov1996}), and assuming no temperature dependence (their $\alpha=0$), and after taking the short wavelength limit by keeping the two highest order $k$ terms, we arrive at exactly our derived Equation \ref{eq:planar_growthrate}. We suspect the differences arise in how the short wavelength limit was taken, as \citet{Schekinov1996} inconsistently took limits strictly in terms of $\Omega$ even in the presence of $k$ terms (where $\Omega = \omega t - k x$), rather than solely $k$.


\bsp	
\label{lastpage}
\end{document}